\documentstyle[preprint,prb,aps]{revtex}
\tightenlines
\begin{document}
\draft
\title{Quantum  cavitation in  liquid $^3$He: dissipation effects}
\author{Dora M. Jezek$^1$,
Mart\'{\i} Pi$^2$, and Manuel Barranco$^2$.}
\address{$^1$Departamento de F\'{\i}sica, Facultad de Ciencias Exactas
y Naturales, \\
Universidad de Buenos Aires, RA-1428 Buenos Aires, and \\
Consejo Nacional de Investigaciones Cient\'{\i}ficas y T\'ecnicas,
Argentine}
\address{$^2$Departament d'Estructura i Constituents de la Mat\`eria,
Facultat de F\'{\i}sica, \\
Universitat de Barcelona, E-08028 Barcelona, Spain}

\date{\today}

\maketitle

\begin{abstract}

 We have investigated the effect that dissipation may have on the
cavitation process in normal liquid  $^3$He.
Our results  indicate that a  rather small dissipation
decreases sizeably the quantum-to-thermal crossover temperature $T^*$
for cavitation in normal liquid $^3$He. This is a possible
explanation why recent experiments have not yet found
 clear evidence of quantum cavitation at
temperatures below the $T^*$ predicted by calculations which neglect
dissipation.

\end{abstract}


Quantum cavitation in superfluid liquid $^4$He has been unambiguously
observed using  ultrasound experimental techniques \cite{Ba95,La98}.
These experiments have shown that quantum cavitation
 takes  over thermal cavitation at a
temperature ($T$) around  200 mK, in good agreement with theoretical
calculations \cite{Ma95,Gu96}, so that  the problem of cavitation in
liquid $^4$He can be  considered as satisfactorily settled.

The crossover temperature corresponding to $^3$He has also been
calculated\cite{Ma95,Gu96}, predicting that $T^* \sim$ 120 mK.
It turns out that preliminary results obtained in a recent experiment
\cite{Ca98} have not shown clear evidence of quantum cavitation  for
 temperatures even below that value. However, the phenomenon has been
firmly established as a stochastic process.
A possible explanation is that  thermal cavitation  is still
the dominant process down to temperatures lower than predicted.

The method  of Ref. \onlinecite{Gu96} (see also Ref. \onlinecite{Je97})
is based, on the one hand, in using a density functional  that
reproduces the thermodynamical
properties of liquid $^3$He at zero temperature (equation of state,
effective mass, etc), as well as the properties of the $^3$He
free surface.  A major
advantage of using a density functional is that one can handle
 bubbles in the vicinity of the spinodal region, where they
are not empty objects \cite{Ma95,Gu96} and
any attempt to describe the critical bubble in terms of a sharp
surface radius fails \cite{Bu96}.
On the other hand, we have used  a functional-integral approach
especially well suited to find $T^*$. This gives us some confidence
on the values obtained for the crossover temperature, and inclines
us to think
that any appreciable discrepancy between theory and experiment has
to be attributed  not to  the method itself, but to some physical
ingredient which has been overlooked in the formalism.
One such  ingredient in the case of liquid $^3$He
is dissipation, which is known\cite{Ha90} to decrease $T^*$.
Since $^4$He is superfluid below the lambda temperature, we
are actually treating both quantum fluids within the same
framework, the behavior of $^4$He being accounted for by the
dissipationless version of the general formalism.

Our starting point is the real time Lagrangian density
${\cal L}(\rho, s)$
\begin{equation}
{\cal L}(\rho, s) =  m \dot{\rho} s - {\cal H}(\rho, s) \,\, ,
\label{eq1}
\end{equation}
where $\rho(\vec{r},t)$ denotes the particle density,  $m$ the
$^3$He atomic mass, and $s(r,t)$ is the velocity potential, i.e,
the  collective velocity  is
$\vec{u}(\vec{r},t) = \nabla s(\vec{r},t)$. The Hamiltonian
density ${\cal H}(\rho, s)$ reads
\begin{equation}
 {\cal H}(\rho, s)
 =\frac{1}{2} m \,\rho\,{\vec u}\,^2
+\left[\omega(\rho)-\omega(\rho_{m})\right]\,\, ,
\label{eq2}
\end{equation}
where $\omega(\rho)$ is the grand potential density of the system and
$\rho_m$ is the density of the metastable homogeneous liquid. We
refer the reader to Ref. \onlinecite{Gu96} and references therein for
details.

To describe the dynamics in the dissipative regime while still being
able to deal with inhomogeneous $^3$He, which is  crucial for
a proper description of cavitation in liquid helium, we have
introduced a phenomenologycal Rayleigh's dissipation
function\cite{Go80,Ga70} ${\cal F}$
\begin{equation}
 {\cal F} =\frac{1}{2} \,\xi \,\frac{{\dot{\rho}}^2}{{\rho}^2}
\,\, .
\label{eq3}
\end{equation}
From Lagrange's equations
\begin{equation}
 \frac{\partial}{\partial \,t} \left(\frac{\delta{\cal L}}{\delta
\dot{x}}\right)  -  \frac{\delta {\cal L}}{\delta x} =
-  \frac{\partial{\cal F}}{\partial \dot{x}}
\,\, ,
\label{eq4}
\end{equation}
with $x$ being either $s$ or $\rho$, one gets the
continuity and  motion equation, respectively:
\begin{equation}
\dot{\rho} + \nabla(\rho \, \vec{u}) = 0
\label{eq5}
\end{equation}
\begin{equation}
m \left\{\frac{\partial \,u_k}{\partial\,t} + u_i \nabla_k u_i
\right\} = - \nabla_k \left(\frac{\delta \omega}{\delta \rho}\right)
+ \nabla_k \left[ \xi \frac{1}{{\rho}^2}
\nabla(\rho\,\vec{u})\right] \,\, .
\label{eq6}
\end{equation}
For an homegeneous fluid, the equation of motion ressembles the
Navier-Stokes equation \cite{La71}
\begin{equation}
m \,\rho \left\{\frac{\partial \,u_k}{\partial\,t} + u_i \nabla_k u_i
\right\} = - \nabla_k P + \eta \Delta u_k +
 \left(\zeta + \frac{1}{3} \,\eta\right) \nabla_k (\nabla\cdot\vec{u})
\,\, ,
\label{eq7}
\end{equation}
where $P$ is the pressure. For liquid $^3$He at low $T$,
dissipation depends on the mean free path of quasiparticles, and a
precise estimation of the magnitude of this effect in the tunneling
process is difficult.
Since our interest here is to explore the effect of a small viscosity
on $T^*$, we have adopted the pragmatic point of view of identifying
$\xi$ with $ \zeta + \eta/3$ and presenting results for different
$\xi$'s close to the experimental $\eta$ value
(it is known that at low temperatures, the shear viscosity coefficient
$\eta$ is much larger than the bulk  viscosity coefficient $\zeta$,
see for example Ref. \onlinecite{Ba91}). Using the macroscopic viscosity
coefficient, one should have in mind that we are likely overestimating
the dissipation effects.

To obtain $T^*$ we have proceeded as indicated in  Ref. \onlinecite{Gu96},
writing the above equations in imaginary time $\tau = i t$ and
linearizing them around the critical bubble density $\rho_0$, seeking
solutions of the kind:
\begin{equation}
\rho(\vec{r},\tau) \equiv \rho_0(r) + \rho^1(r)\, e^{- i \omega_s
\tau}
\, \, .
\label{eq8}
\end{equation}
Upon linearization, we end up with the following equation for $\omega_s$
and $\rho^1(r)$:
\begin{equation}
{\cal M} \rho^1(r)  \equiv
\left[m \,\omega_s^2 - {\cal M}_1 - \xi \omega_s {\cal M}_2 \right]
\rho^1(r)  = 0
\,\, .
\label{eq9}
\end{equation}
The differential operators ${\cal M}_1$ and ${\cal M}_2$
in Eq. (\ref{eq9}) are, respectively, the linearization of
\begin{equation}
\nabla\left\{\rho \nabla\left( \frac{\delta \omega}{\delta \rho}
\right)\right\} \,\,\, {\rm and} \,\,\,
\nabla \left\{\rho \nabla\left(\frac{1}{\rho^2} \right)
\right\}
\, ,
\label{eq10}
\end{equation}
in which only first order terms in $\rho^1(r)$ and its derivatives
have been kept \cite{Gu96}.
Since $\xi$ depends on the density\cite{Ba91} as $\rho^{5/3}$,
in actual calculations we have made a local density approximation,
using as form factor in Eq. (\ref{eq3})
the expression $1/(\rho_{sat}^{5/3} \rho^{1/3}(r))$, where $\rho_{sat}$
is the density of the liquid at $T$ = 0 and $P$ = 0,
and $\xi$ is then  density-independent.

Eq. (\ref{eq9}) is a fourth-order linear differential,
generalized eigenvalue equation, whose physical solutions have to
fulfill $\rho^{1\prime}(0) = \rho^{1\prime\prime\prime}(0) = 0$, and
fall
exponentially to zero at large distances. We have solved it as indicated
in Ref. \onlinecite{Gu96}. Once the largest dissipation-renormalized
frequency $\omega_s$ has been determined\cite{note},
the crossover temperature is obtained as
$T^* = \hbar \omega_s/(2 \pi)$.

Table I collects the equation of state near the spinodal point
($\rho_{sp}$ = 0.01191 ${\rm \AA}^{-3}$, $P_{sp} = -3.102$ bar),
 and other quantities
which are of interest to analyze the experimental
results \cite{La98}. Our spinodal point compares very well with recent
Monte Carlo calculations\cite{Bo98}
($\rho_{sp}$ = 0.0121 ${\rm \AA}^{-3}$, $P_{sp} = -3.12\pm0.10$ bar),
and also with other phenomenological approaches \cite{Ma95,So92}.

We show $T^*$ in Fig. \ref{fig1} as a function of pressure for
different $\xi$ values. In particular, $\xi$ = 100 $\mu$P
roughly corresponds to the experimental value\cite{Pa78} of $\eta$ at
$P$ = 0 and $T$ = 100 mK. The associated  effective quantum
action ${\cal S}$ obtained as  ${\cal S} = \Delta \Omega / T^*$, where
$\Delta \Omega$ is the maximum of the energy barrier,
is displayed in Fig. \ref{fig2}.

Fig. \ref{fig3} shows  $\rho^1(r)$  at $P = -3$ bar for
three $\xi$ values, as well as the critical bubble density $\rho_0(r)$.
The linearized continuity equation
$\rho^1(r) \propto \nabla(\rho_0 \, \vec{u})$ implies that $\rho^1(r)$
must have nodes, as it imposes that the integral of $\rho^1(r)$ is zero
when taken over the whole space.

When $\xi$ is small enough and the ${\cal M}_2$ term in Eq. (\ref{eq9})
can be treated perturbatively, a straightforward calculation yields
\begin{equation}
\omega_s =  \sqrt{\omega_{0,0}^2 +
\left(\frac{\xi \mu_{\,2}}{2\,m}\right)^2}
- \frac{\xi \mu_{\,2}}{2\,m}
\,\, ,
\label{eq11}
\end{equation}
where we have used a standard matrix notation \cite{Go80,Ga70} to
denote as $\omega_{0,0}$ and $|\rho^{1\,(0)}_0\rangle$
 the higher frequency solution of the non-viscous problem
$(m\, \omega^2_{0,n} - {\cal M}_1)
|\rho^{1\,(0)}_n\rangle$ = 0, and have defined
 $\mu_{\,2} \equiv -
\langle \rho^{1\,(0)}_0 | {\cal M}_2 |\rho^{1\,(0)}_0\rangle > 0$.
Equation (\ref{eq11}) is similar to that given in Ref. \onlinecite{Ha90}
for the dissipation-renormalized frequency $\omega_s$ in the case of
frequency-independent damping.

Figures (\ref{fig1}-\ref{fig2}) indicate that for viscosity values of
the order of the experimental one, a sizeable decrease of the
crossover temperature occurs. However, the present model still predicts
that a transition from thermal to quantum
cavitation takes place in liquid $^3$He.

We finally obtain the homogeneous
cavitation pressure $P_h$ from the equation \cite{Gu96}:
\begin{equation}
1 = (V t)_{exp}\, J_0\, e^{-{\cal S}} \,\, ,
\label{eq14}
\end{equation}
taking for the experimental volume$\times$time $(V t)_{exp}$
a typical value of $10^8\, {\rm \AA}^3$ s, which
correspods to $^4$He experiments \cite{La98}.
We have adopted for $J_0$ the same prescription as in Ref.
\onlinecite{Gu96}.
Figure \ref{fig4} shows the homogeneous cavitation pressure as a
function of $T$ for the $\xi$ values we have been using.

In conclusion, we have developed a phenomenological model to size
the effect of dissipation in the cavitation process
in liquid $^3$He that allows one to handle realistic critical
cavitation configurations near the spinodal line, and
to treat both helium isotopes within the same frame, using the
dissipationless limit of the method in the case of $^4$He.
The results we have obtained indicate that for liquid $^3$He
even a moderate dissipation may reduce the crossover temperature
in a non-negligible amount, displacing the homogeneous cavitation
pressure towards the spinodal value.
Viscosity may then be the reason of the inconclusive
results for quantum cavitation reported in Ref.
\onlinecite{Ca98} which, if confirmed, would indicate that dissipation
plays a crucial role in quantum cavitation in liquid helium.
The experimental study of cavitation in undersaturated
$^3$He-$^4$He mixtures might then uncover a structure much richer than
that theoretically described in Ref. \onlinecite{Je97}, since $^4$He is
still superfluid
and $^3$He is in the normal phase. This would open the possibility of
studying the  influence of dissipation in
the cavitation process varying the $^3$He concentration.

We would like to thank  Sebastien Balibar, Eugene Chudnovsky
and Patrick Roche for useful discussions.
This work has been supported in part by  the DGICYT (Spain),
grant PB95-1249, and by the Generalitat de Catalunya `Accions
Integrades' and 1998SGR-00011 programs.

%
%
\begin{table}
\caption{Equation of state, sound velocity, energy barrier and
quantum action ($\xi$ = 0) near the spinodal point.}
\begin{tabular}{ccccc}
   $\rho$ &  $P$ &  $c_s$  &  $\Delta\Omega$  & ${\cal S}/\hbar$ \\
(${\rm \AA}^{-3}$) & (bar) &  (m/s) &     (K)   &
 \\ \tableline
     .0123  &   -3.08 &   42.3    &      1.3    &     16.0    \\
     .0124  &   -3.07 &   47.8    &      1.8    &     19.4    \\
     .0125  &   -3.06 &   52.7    &      2.5    &     22.9    \\
     .0126  &   -3.05 &   57.4    &      3.2    &     26.8    \\
     .0127  &   -3.03 &   61.8    &      4.0    &     31.2    \\
     .0128  &   -3.01 &   66.1    &      4.9    &     36.0    \\
     .0129  &   -2.99 &   70.1    &      5.8    &     41.5    \\
     .0130  &   -2.96 &   74.1    &      6.9    &     47.8    \\
     .0131  &   -2.93 &   77.9    &      8.0    &     55.0    \\
     .0132  &   -2.90 &   81.7    &      9.3    &     63.5    \\
     .0133  &   -2.86 &   85.3    &     10.6    &     73.5    \\
     .0134  &   -2.83 &   88.9    &     12.1    &     85.5    \\
     .0135  &   -2.78 &   92.4    &     13.7    &    100.3    \\
     .0136  &   -2.74 &   95.9    &     15.5    &    119.2    \\
     .0137  &   -2.69 &   99.3    &     17.4    &    144.0    \\
     .0138  &   -2.64 &  102.7    &     19.6    &    175.4    \\
     .0139  &   -2.59 &  106.1    &     21.9    &    212.6    \\
     .0140  &   -2.53 &  109.4    &     24.6    &    256.6    \\
\\
\end{tabular}
\end{table}
\begin{figure}
\caption{  $T^*$ as a function of pressure for different $\xi$
values (in $\mu$P). The homogeneous cavitation pressure
 $P_h(T^*)$ is shown as circles  for  $(V t)_{exp} = 10^8\,
 {\rm \AA}^3$ s.}
\label{fig1}
\end{figure}
\begin{figure}
\caption{ Effective quantum action in $\hbar$ units as a function of
pressure for the values of $\xi$ employed in Fig. 1.}
\label{fig2}
\end{figure}
\begin{figure}
\caption{ Particle density profile $\rho_0(r)$ in ${\rm \AA}^{-3}$
for $P = -3$ bar (solid
line), as well as $\rho^1(r)$ densities for three $\xi$ values
(arbitrary units).}
\label{fig3}
\end{figure}
\begin{figure}
\caption{ Homogeneous cavitation pressure  $P_h$  as a function of $T$
for $(V t)_{exp} = 10^8 \, {\rm \AA}^3$ s, and
 the $\xi$ values employed in Figs. 1 and 2.}
\label{fig4}
\end{figure}

\begin{references}
\bibitem{note} In actual calculations, we have only found one physically
acceptable solution, as in the non-viscous case, see Ref. \onlinecite{Gu96}.
\end{references}
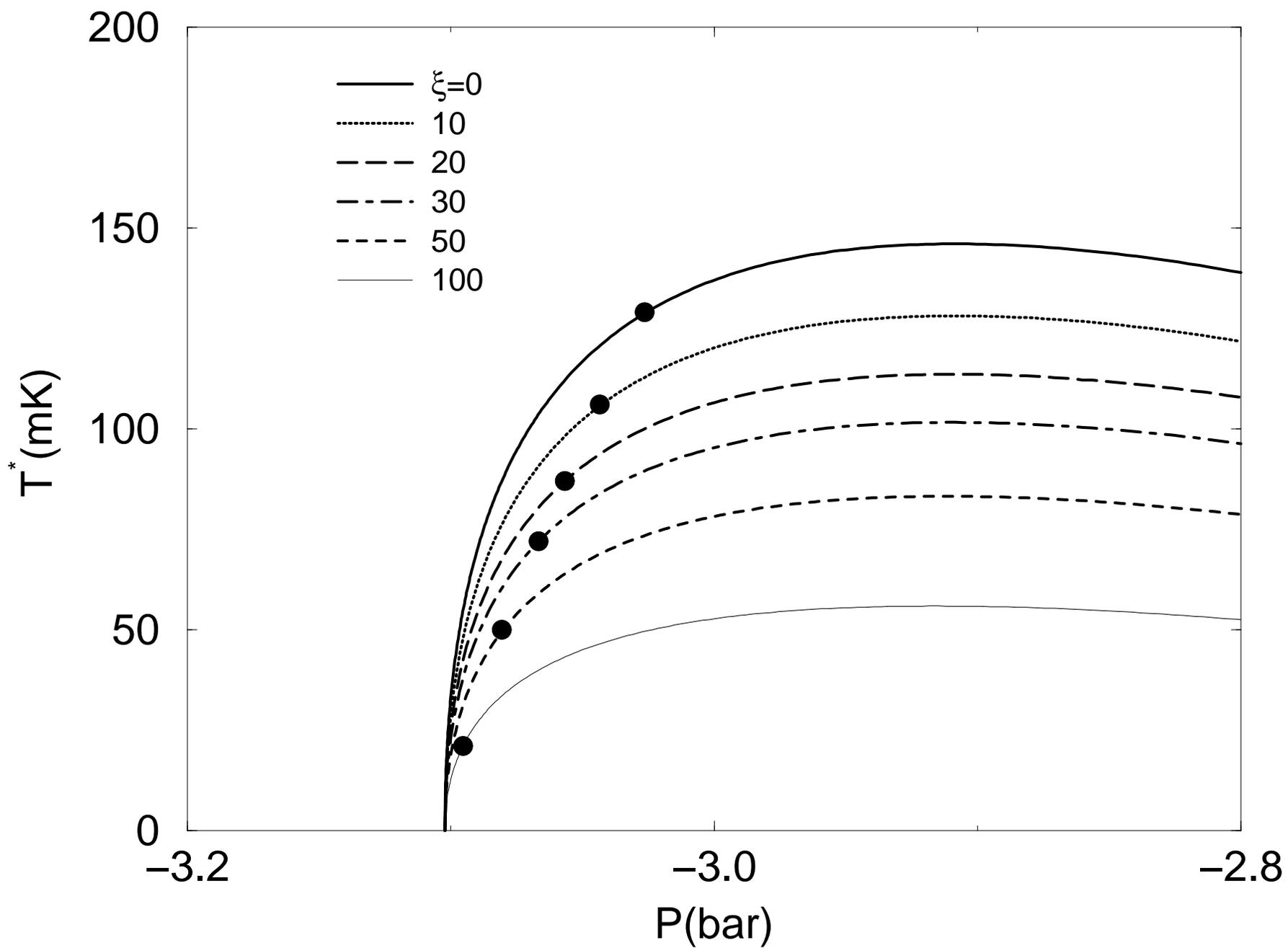
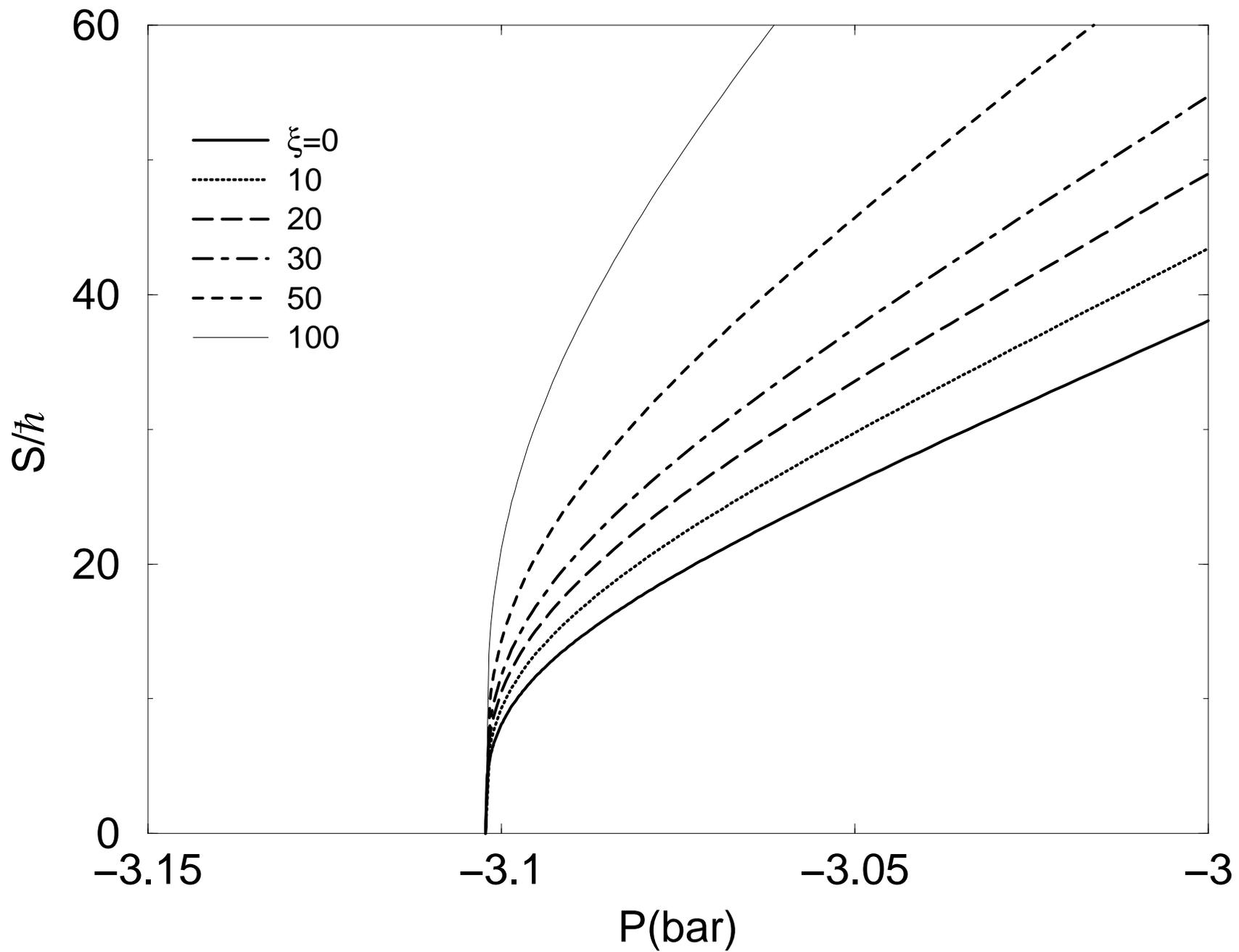
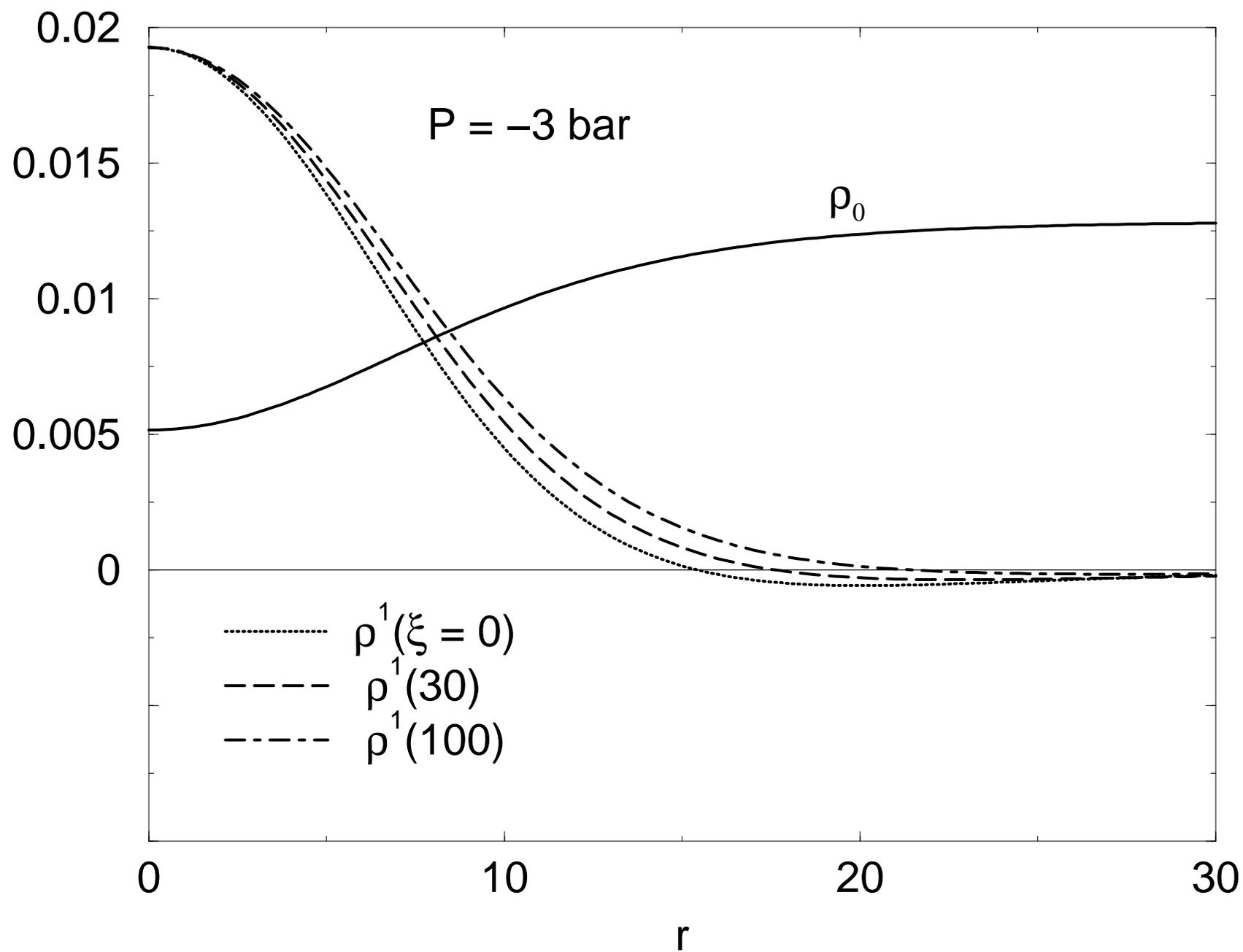
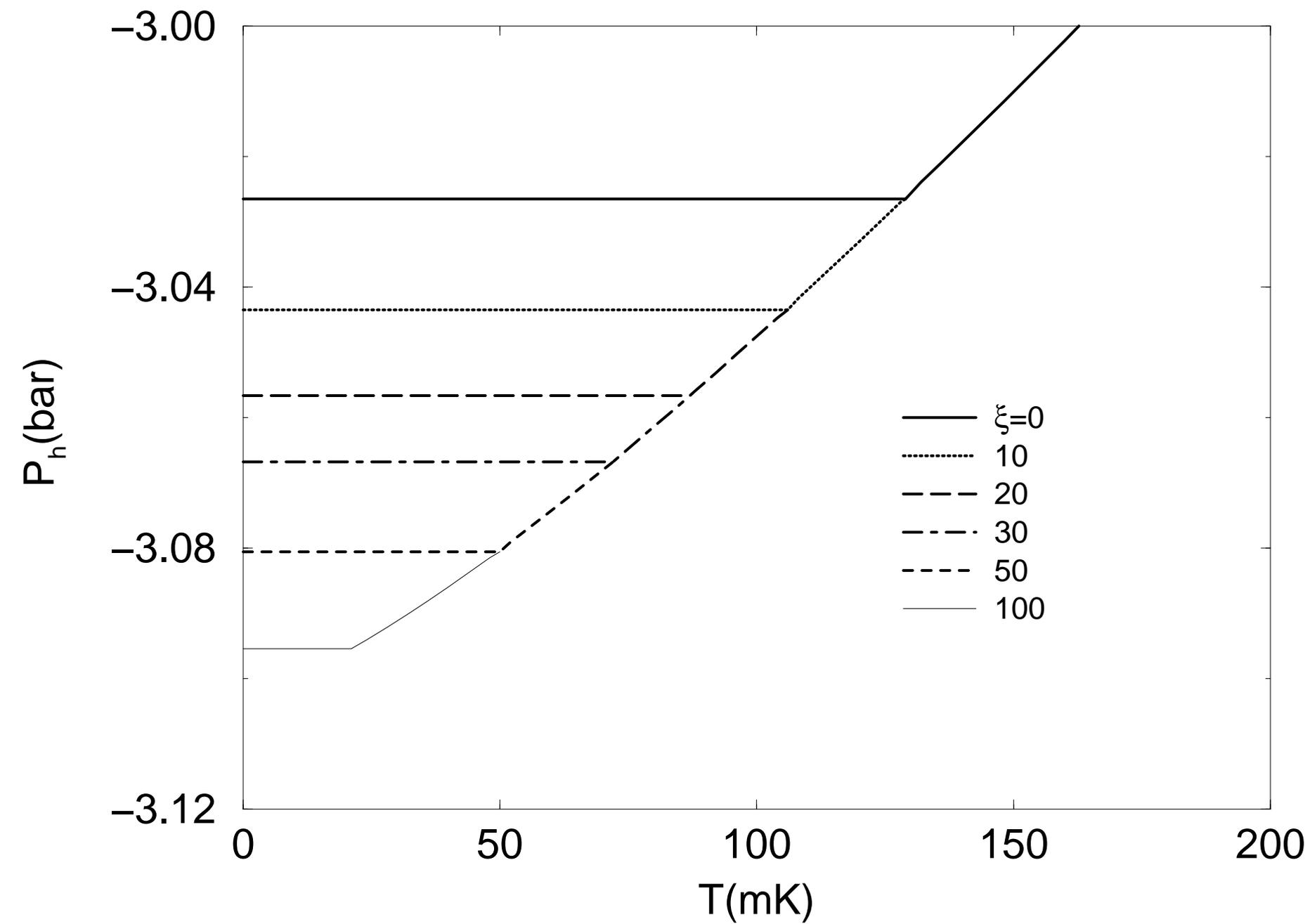
\end{document}